# Valley splitting in strained silicon quantum wells

Timothy B. Boykin[1], Gerhard Klimeck[2], M. A. Eriksson[3], Mark Friesen[3,4],
S. N. Coppersmith[3], Paul von Allmen[2], Fabiano Oyafuso[2], and Seungwon Lee[2]

*[1] Department of Electrical and Computer Engineering,
The University of Alabama in Huntsville, Huntsville, Alabama 35899*
*[2]Jet Propulsion Laboratory, California Institute of Technology,
4800 Oak Grove Road, MS 169-315, Pasadena, California 91109*
*[3]Department of Physics, The University of Wisconsin, Madison, Wisconsin 53706*
*[4]Department of Material Science and Engineering, The University of Wisconsin,
Madison, Wisconsin 53706*

A theory based on localized-orbital approaches is developed to describe the valley splitting observed in silicon quantum wells. The theory is appropriate in the limit of low electron density and relevant for proposed quantum computing architectures. The valley splitting is computed for realistic devices using the quantitative nanoelectronic modeling tool NEMO. A simple, analytically solvable tight-binding model is developed, it yields much physical insight, and it reproduces the behavior of the splitting in the NEMO results. The splitting is in general nonzero even in the absence of electric field in contrast to previous works. The splitting in a square well oscillates as a function of $S$, the number of layers in the quantum well, with a period that is determined by the location of the valley minimum in the Brillouin zone. The envelope of the splitting decays as $S^3$. Finally the feasibility of observing such oscillations experimentally in modern Si/SiGe heterostructures is discussed.

Recently there has been a great deal of interest in developing semiconducting nanostructures in which spins are coherent; one example is a proposed spin-based quantum dot quantum computer [1]. Reasons for using heterostructures made of silicon as opposed to gallium arsenide include (1) longer intrinsic spin coherence times due to smaller spin-orbit coupling and (2) elimination of decoherence caused by coupling between electrons and nuclear spins by use of commercially available isotropically pure spin-zero $^{28}$Si[2]. However, one complication of Si compared to GaAs is that unstrained Si has a six-fold degenerate conduction-band minimum. Strain in Si/SiGe heterostructures reduces the six-fold valley degeneracy to a two-fold one, but the remaining two-fold valley degeneracy is a potential source of decoherence and other difficulties [3]. It is thus of great interest to understand how to lift this remaining two-valley degeneracy and maximize the splitting between these two levels.

Valley splitting in Si heterostructures has been studied both experimentally [4] and theoretically for many years; ref. [5] provides an exhaustive review of many efforts. Early work includes the effective mass approaches of Sham and Nakayama [6] at single interfaces, and Ohkawa [7] in quantum wells. More recently, Grosso, et al. [8] use an approach based on an $sp^3$ empirical tight-binding model to study the splitting in Si superlattices. None of these works focus on the essential differences in the behavior of the valley splitting in triangular potentials and in square wells, and with the development of modulation-doped heterostructures, square well potentials are much more relevant experimentally than previously. Interestingly, Ohkawa [7] finds the essential features of the valley splitting in finite quantum square wells: the presence of a nonzero splitting even in zero electric field; the oscillation of the splitting as a function of quantum well thickness; and the overall decay of the splitting as the cube of the well thickness. However, Ohkawa did not appreciate the essential role of the square well potential in obtaining these results. Moreover, Ohkawa's methods have been criticized [5]; an important conceptual problem is that his multi-band effective mass theory uses a *k*-state basis, fundamentally inappropriate for a quantum-confined structure, and necessarily includes an intervalley-coupling constant.

Here, the problem of the intervalley coupling in the limit of low electron density is addressed using the empirical tight-binding method. The localized-orbital basis used in tight binding is most appropriate for the conditions prevailing in quantum-confined heterostructures: non-conservation of wavevector, *k*, and large changes of the potential on the atomic length scale. The tight-binding approach has the further advantage that moving from the localized-orbital description to one based upon a superposition of bulk-like states is straightforward and transparent. In contrast, it is not generally possible within the multi-valley effective mass approach to transition from the bulk description of the state to a localized-orbital description.

Fig. 1 shows the valley splitting versus well width at various electric fields for strained Si [001] quantum wells with hard wall (infinite barrier) boundary conditions. The strain conditions are such that a $Si_{0.8}Ge_{0.2}$ relaxed substrate is assumed. The results were obtained numerically using the NanoElectronic MOdeling tool, NEMO [9]. NEMO 1-D is an advanced heterostructure modeling tool designed to simulate high bias transport across heterostructure layers using the non-equilibrium Green function formalism. NEMO 1-D is capable of quantitative electron transport simulation in realistically sized quantum devices built with arbitrary semiconducting materials. NEMO 1-D has served [9] as a quantitatively predictive design and analysis tool for resonant tunneling diodes, oxide thickness analysis for MOS transistors as well as studies of incoherent scattering. In these particular simulations a nearest-neighbor, tight-binding, spin-orbit $sp^3d^5s^*$ model [10] with 40 orbitals per unit cell, recently updated to accommodate strained silicon-germanium heterostructures is being used. NEMO modeling is expected to provide a highly accurate picture of the atomic-scale structure of electronic wavefunctions in a silicon quantum well in the limit of low electron densities appropriate for quantum computing. Three features of the calculation stand out: (*i*) the splitting is nonzero even at zero field; (*ii*) the zero field splitting exhibits oscillations as a function of well width; and (iii) the envelope decays as the cube of the well thickness.

While the NEMO calculation gives results of high accuracy, the sheer number of bands included in the model often makes it difficult to identify the underlying physics. To overcome such difficulty and to gain more insight simpler models which have indirect conduction-band minima are developed. All are essentially the same, being related to one another by basis changes; they are depicted in the left panel of Fig. 2. The mathematically simplest version (top) consists of a chain of cells (length $a/2$) of one atom each, with one $p_z$-like orbital per atom. There are three Hamiltonian parameters: onsite, $\varepsilon$; nearest-neighbor, $v$; and second-near-neighbor, $u$. The resulting bulk (cyclic boundary) model thus yields a single-band. For a quantum well of $S$ atoms with hardwall boundary conditions, the $S \times S$ Hamiltonian matrix is penta-diagonal, with $\varepsilon$ on the diagonal, $v$ on the nearest sub-diagonals, and $u$ on the second-nearest sub-diagonals.

Grouping two atoms per unit cell (length $a$) yields a two-band tight-binding model (middle). Under this trivial change of basis the quantum well Hamiltonian is of course unchanged. (The bulk dispersion now consists of two bands, but in a Brillouin zone half as large as that of the one-band model.) A further basis change consists of taking symmetric and antisymmetric combinations of the orbitals in a cell of the first two-band model (bottom). This yields a two-band model having one atom per cell (length *a*) and two orbitals (pseudo-*S* and pseudo-*P*) per atom. This model makes comparison of wavefunctions with those of NEMO easiest; its bulk dispersions are the same as with the previous two-band model. (Further details will be discussed elsewhere [11].) Because all of these basis transformations are unitary, the quantum well Hamiltonian eigenvalues are unaltered.

The bulk (cyclic boundary) version of the tight-binding model has the dispersion relations

$$E_{\pm}(k) = \varepsilon \pm 2v\cos\left(\frac{ka}{2}\right) + 2u\cos(ka) \quad , \tag{1}$$

where $a$ = 2.715 Å is the length of a unit cell. Adjusting $v$ and $u$ so that the location of the degenerate valley minima and the band curvature near the conduction valley minimum agree with those of the bulk dispersion provided by NEMO yields $v$ = 0.682640 eV and $u$ = 0.611705 eV. The energy difference $E_{21}$ between the two low-lying eigenstates in the finite two-band model is now calculated without additional fitting parameters. The results, shown in Fig. 3, show excellent correspondence with the more sophisticated NEMO model, reproducing the oscillations as well as the functional form of the envelope of the amplitude. Since NEMO incorporates spin-orbit coupling while the two-band model does not, it is clear that (*i*) spin-orbit coupling is not crucial [12], and (*ii*) the simpler two-band model does indeed capture the essential physics of valley splitting.

The following paragraphs will discuss some calculation details, describe an analytical expression for the valley splitting, and provide a physical picture of valley splitting. The lowest conduction band of Si along [001] is plotted in the right panel of figure 2. A constant-energy slice at an energy lying between the minimum of the lowest conduction-band and the degeneracy point at the zone edge intersects *two* different wavevectors $k_1 = k_{\min} + \delta_1$ and $k_2 = k_{\min} - \delta_2$, where $k_{\min}$ corresponds to the minimum point of the conduction band edge along the Δ-line. Thus, there are four Bloch states at this energy $\pm k_1, \pm k_2$, all of which in principle contribute to the total wavefunction when quantum confinement is imposed. While the details of the phase matching at the boundaries are essential for determining the splitting and interpreting the wavefunctions, the physical picture is clear. In a quantum well of length *L* the envelope of each of the two lowest states (*j*=1, 2) is essentially cosine-like, characterized by a wavevector $\bar{\delta}^{(j)} = \left(\delta_1^{(j)} + \delta_2^{(j)}\right)/2 \approx \pi/L$, while the underlying fast oscillations (about $\pi/2$ out of phase relative to one another) are roughly characterized by $k_{\min}$. There are two states with similar envelopes because a conduction band minimum occurring inward from the zone edge supports two states at slightly different energies which nevertheless have approximately the same $\bar{\delta}$.

In carrying out the calculation, the wavefunction is written in the localized-orbital basis for a chain of 2*N*+1 unit cells (each unit cell of the chain represents a monolayer in [001]-oriented Si) centered at the origin, as:

$$|\psi\rangle = \sum_{j=-N}^{+N} \left[ C_j^{(\alpha)} |\alpha; ja\rangle + C_j^{(\beta)} |\beta; ja\rangle \right] \qquad (2)$$

where $\alpha$ and $\beta$ are the localized orbitals. This quantum well is inversion-symmetric, so the eigenstates of the Hamiltonian may be chosen to be simultaneous eigenstates of parity. The energies of interest for valley-splitting belong to the valley of the lower band with only two Bloch states $k_1$ and $k_2$, satisfying Eq. (1),

$$E_-(k_1) = E_-(k_2) \tag{3}$$

The localized-orbital expansion coefficients, $C$, can therefore be expressed in terms of these two states [13]. For the even envelope states the $C$ are linear combinations of cosines at $k_1$ and $k_2$, while for the odd envelope states they are combinations of sines. The hard wall condition requires that:

$$C_{N+1}^{(\alpha)} = C_{N+1}^{(\beta)} = 0, \quad C_{-(N+1)}^{(\alpha)} = C_{-(N+1)}^{(\beta)} = 0\,. \tag{4}$$

Eqs. (3) and (4) yield a system of two simultaneous equations that must be satisfied. The equations can be solved analytically order-by-order in powers of $(S+2)^{-1}$, where $S = 2(2N+1)$, for the two lowest bound states [11]. To leading order in $(S+2)^{-1}$, the splitting between these levels, denoted $E_{21}$, is

$$E_{21} \approx \frac{16\pi^2 u}{(S+2)^3}\left|\sin\left[(S+2)\phi_0\right]\right|\sin(\phi_0), \quad \sin(\phi_0) = \sqrt{1-\left(\frac{v}{4u}\right)^2} \tag{5}$$

where $\phi_0 = k_{\min}a/2$, and $k_{\min}$ is the wavevector of the valley minimum. Higher-order corrections to Eq. (5) can be calculated in a straightforward manner [11]; in practice, these corrections are quite modest down to quantum well widths of order 40 unit cells.

Eq. (5), predicts a decay in the amplitude of the oscillations with a $(S+2)^{-3}$ pre-factor. The oscillations with well width have a frequency determined by the location of the valley minimum $k_{\min}$, and are a direct consequence of the phase matching at the interface becoming almost identical for the two lowest states. A corollary of this oscillatory behavior is that the parity of the ground state alternates between even and odd (although both states of the doublet have very similar, cosine-like envelopes). The alternating parity has been noted before [14], but has not been explained.

In Fig. 3, valley splitting results are shown for both NEMO and the 2-band tight-binding model. The expression for $E_{21}(S)$ in Eq. (5) is scaled in amplitude [15] so that it falls on top of NEMO data points (solid curve). The slight quantitative difference of the two models is not surprising, given their different orbital makeup. Because of the many bands included in NEMO, there will be numerous evanescent states present in its wavefunction that are not present in the two-band model. However, the predicted period, phase, and decay of the oscillations agree well.

Although both the oscillations as a function of well length and the decay in their amplitude have been calculated in Ohkawa's multi-valley effective-mass model[7], there are several important differences between the present result and his. Most notably, Ohkawa[7] finds that the splitting is proportional to the band gap at Γ and to the inverse cube of the valley minimum wavevector, unlike eq. (5). It should be remembered that the qualitative similarities (oscillations and their decay) are characteristic of any model based upon two degenerate, interacting Bloch states of different wavevector when the splitting is calculated to lowest surviving order in the inverse quantum well length.

All considerations described above are based on the assumption of a flat band quantum well. However, local electric fields are ubiquitous in heterostructures due to modulation doping and the need for external gate potentials; these effects are addressed in Figs. 1 and 4. Figure 1 displays the splitting versus well width for different values of constant electric field, calculated using NEMO, and allows one to determine whether oscillations as a function of quantum well thickness can be observed experimentally. The field has little effect until the voltage drop per unit cell is of the same order as the splitting at zero field. For a fixed field the oscillations are quenched for longer wells; this result is reasonable since in longer wells the states are more readily localized in the bottom of the notch, where they are insensitive to the location of the far boundary. Figure 4 shows the splitting versus applied field for quantum wells of various length, also calculated using NEMO. As seen in Fig. 4, the splitting for a fixed length increases monotonically as a function of field, becoming linear for higher fields, in agreement with Sham and Nakayama's [6] result for semi-infinite systems. However, at lower fields the splitting in a quantum well is markedly nonlinear, in contrast to that in the semi-infinite system[6].

All discussions above considered a treatment of an infinite hard wall confinement. The effect of a finite voltage discontinuity at the well edges have also been investigated using both NEMO and using the two-band model. The behavior is qualitatively unaffected down to band offsets of order a few tenths of an eV, so the results obtained for infinite square wells should also be realistic guide to the behavior of actual heterostructures [16].

Finally the results presented here need to be related to experimental measurements of valley splitting in Si quantum wells. Several groups have measured nonzero valley splittings with magnitude of the same order as predicted by our models [4]; in the past, these splittings have been usually interpreted as resulting from nonzero electric fields that are typical in modulation-doped heterostructures [6-7]. Indeed, the electric fields from the dopants at typical electron densities ($10^{11}/cm^2$) are indeed such that the voltage drop per unit cell is the same order of magnitudes as the observed splittings, which in turn are of the same order as the zero-field splitting calculated here at well widths of about 10 nm. Lowering the electron density by an order of magnitude will reduce the electric field and suppress many-body effects. The simulations and the model presented here predict that experiments on heterostructures with lower electron density will provide unambiguous evidence for the mechanism for zero-field valley splitting investigated here.

In conclusion, tight-binding calculations, which explain the valley splitting in Si quantum-confined heterostructures are presented. NEMO multiband calculations [9] give the quantitative details while two-band calculations elucidate the physics of these structures. In particular, zero-field splitting oscillations with well width are predicted and explained, reasons for the amplitude decay of the oscillation and reasons for the alternating parity of the ground state are given. The results lead to a better understanding of these important nanostructures.

We acknowledge useful conversations with R. Joynt. Work at JPL, UAH, and UW was sponsored in major proportion by the U. S. Army Research Office through the ARDA program and directly through ARDA. The work at UW was also supported by the National Science Foundation through the MRSEC and QuBIC programs. Part of the work described in this publication was carried out at the Jet Propulsion Laboratory, California Institute of Technology under a contract with the National Aeronautics and Space Administration. Funding was provided at JPL under grants from ARDA, ONR, and JPL.

## Figure Captions

**Figure 1:** Valley splitting versus well width at various applied fields for a strained Si quantum well with hardwall boundary conditions, calculated using NEMO's nearest-neighbor spin-orbit $sp^3d^5s^*$ model. Although calculations are only for integral numbers of monolayers, lines are used as a guide to the eye.

**Figure 2:** *Left*: Sketch of different versions of a simple tight-binding model, all related by basis transformations. Dotted regions are negative, striped regions positive. Top: Single-band model with one *p*-like orbital per atom and one atom per unit cell (length $a/2$). Parameters are $\varepsilon$ (onsite), $v$ (nearest-neighbor), and $u$ (second-near-neighbor). Middle: By grouping two atoms together in a single cell we obtain a two-band model with two atoms per unit cell (length *a*), and each atom with one *p*-like orbital. Bottom: A basis transformation on the middle model using symmetric and antisymmetric combinations yields a two-band model with one atom of two orbitals per unit cell (length $a$). The parameters are: $\varepsilon_s = \varepsilon - v, \varepsilon_p = \varepsilon + v$, $V_{ss} = u - v/2, V_{pp} = u + v/2, V_{sp} = v/2$. *Right*: Lowest conduction bands of strained Si as reproduced by the NEMO nearest-neighbor spin-orbit $sp^3d^5s^*$ model (solid, light curve) and the lowest two conduction bands of a two-band, second-near-neighbor model (no spin-orbit), with parameters given in the text (dark dotted and dashed curves).

**Figure 3:** Valley splitting in a strained Si quantum well at zero applied field with hardwall boundary conditions. Numerical results calculated with NEMO's $sp^3d^5s^*$

model, NEMO-spds, and the two-band model presented here, 2 Band (Exact), are shown as symbols with no lines. The approximate 2 Band splitting from eq. (5), 2 Band (Approx), along with a fit of the NEMO results to eq. (5), spds-fit, are shown as lines with no symbols, although they are calculated at the same points as the numerical results.

**Figure 4:** Valley splitting in a strained Si quantum well with hardwall boundary conditions versus applied field for several well widths in ML (monolayers), as calculated with NEMO's $sp^3d^5s^*$ model. Actual points calculated are shown as symbols.

## Figures

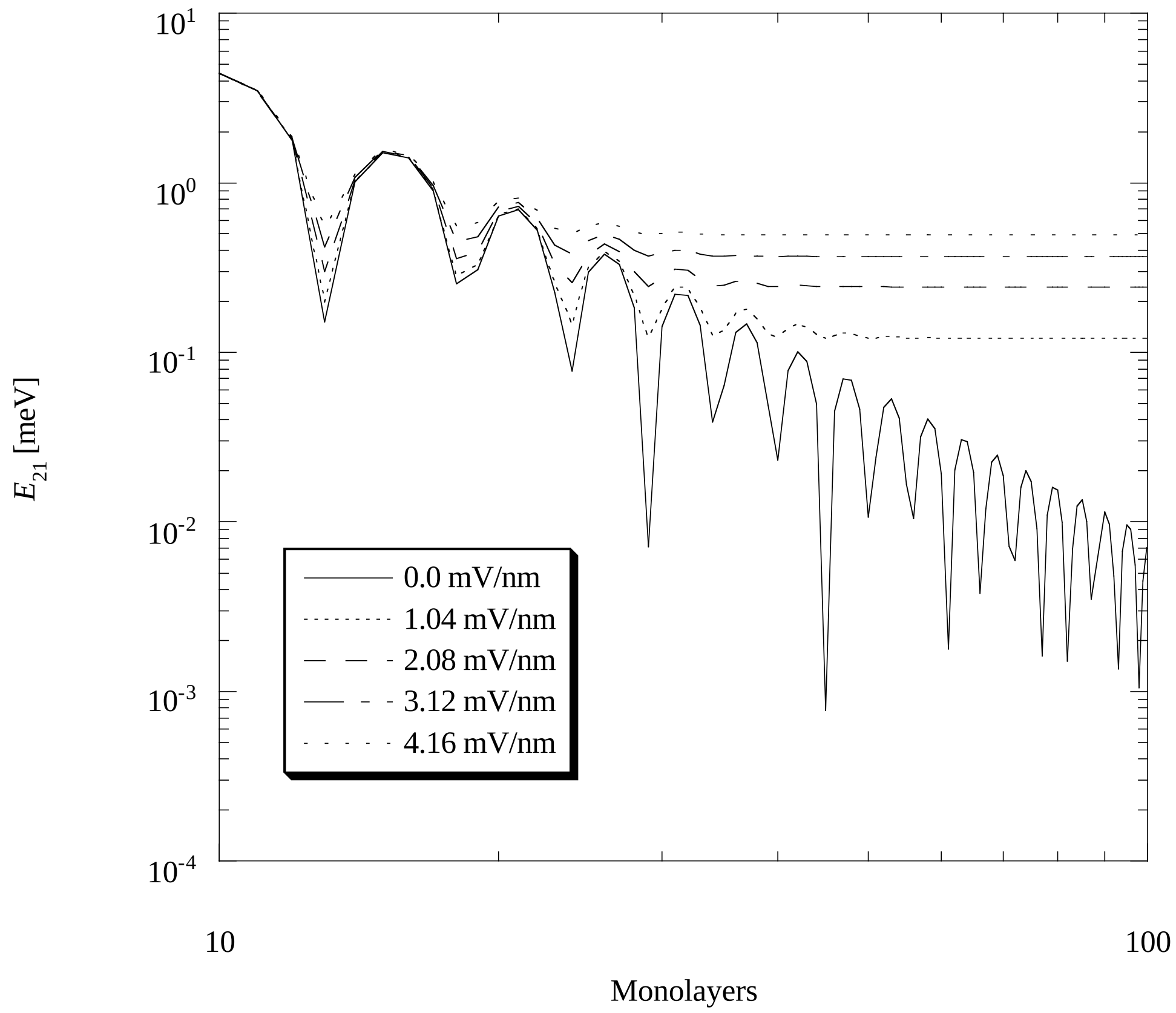

Figure 1

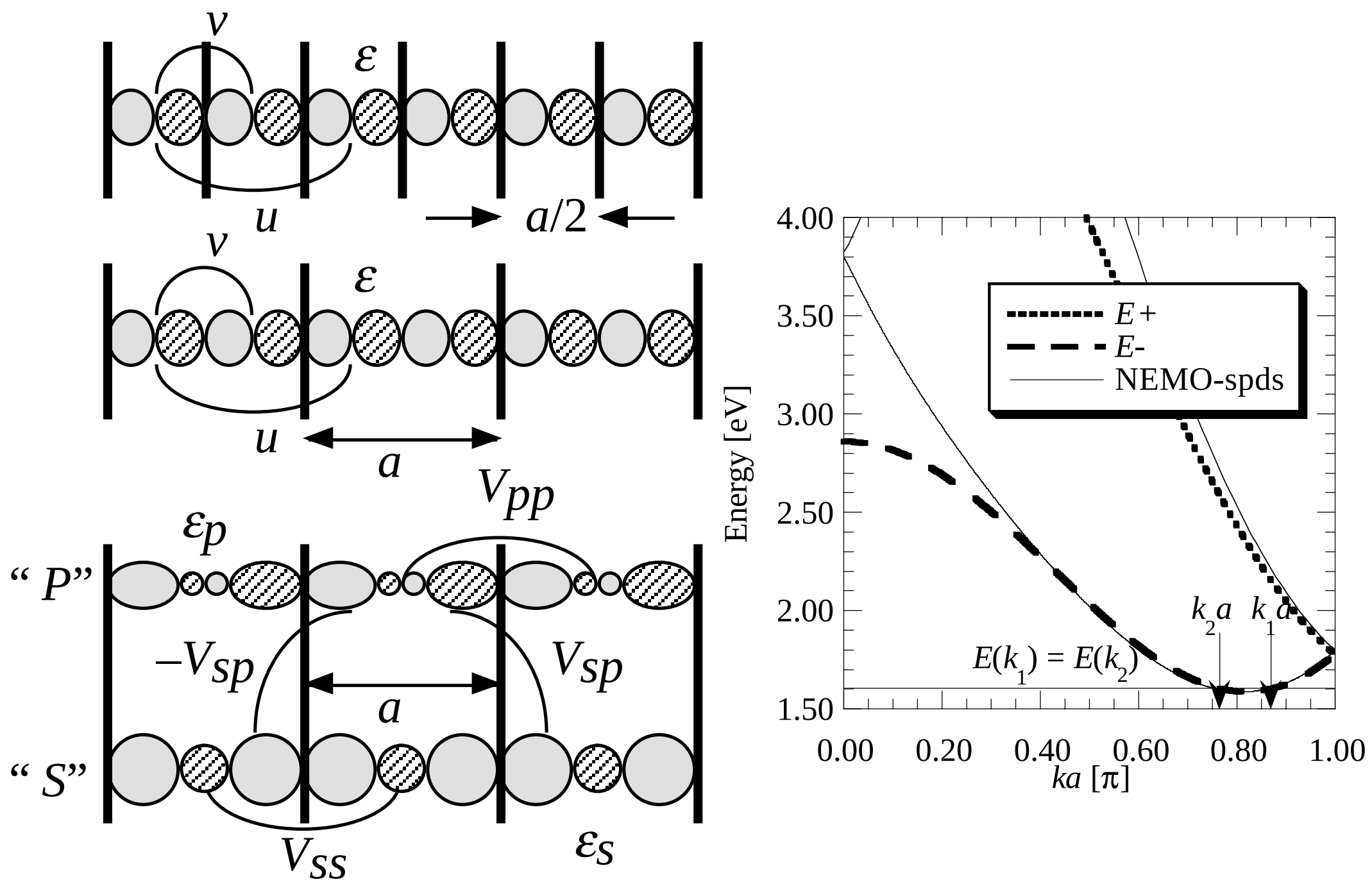

Figure 2

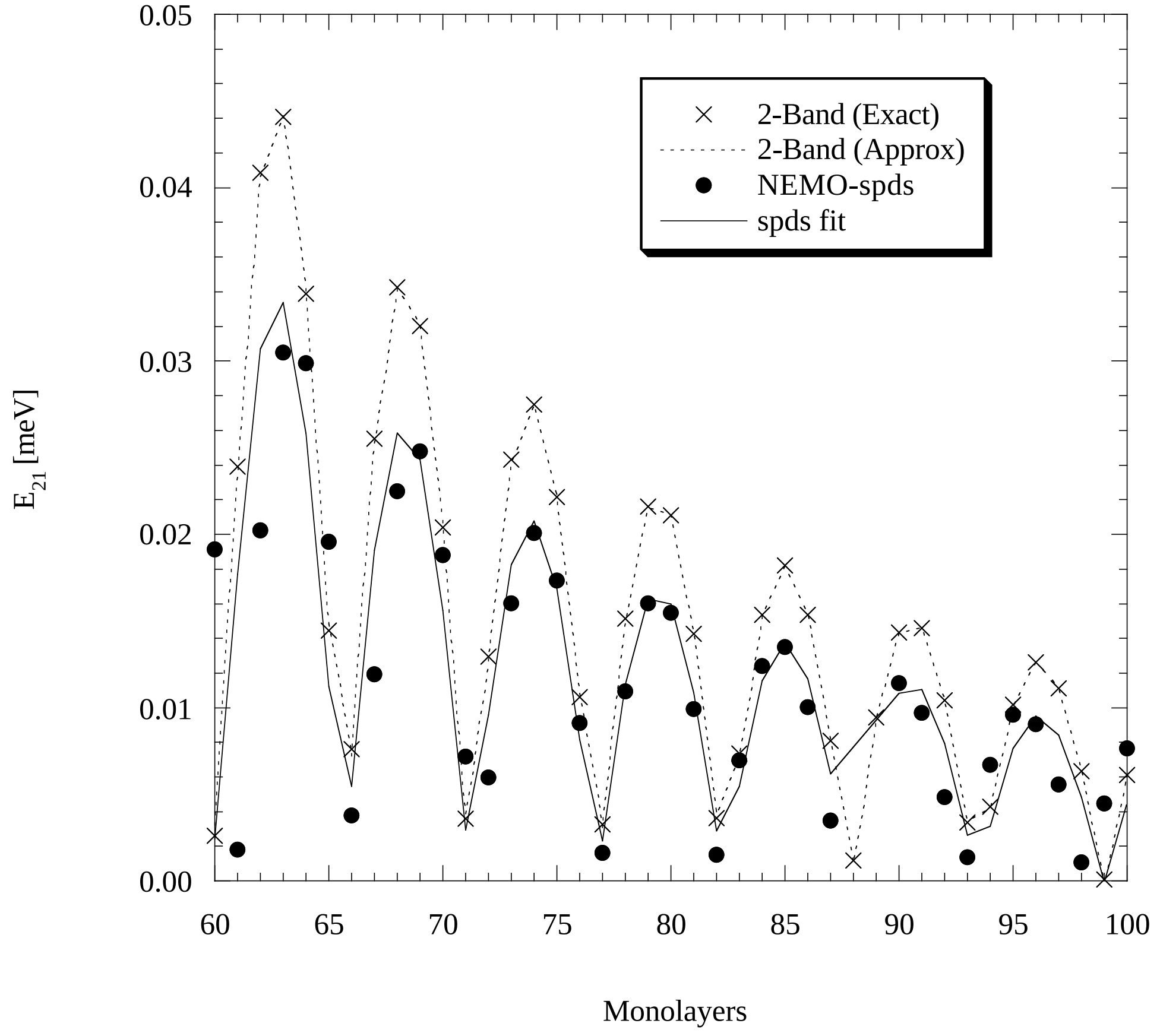

Figure 3

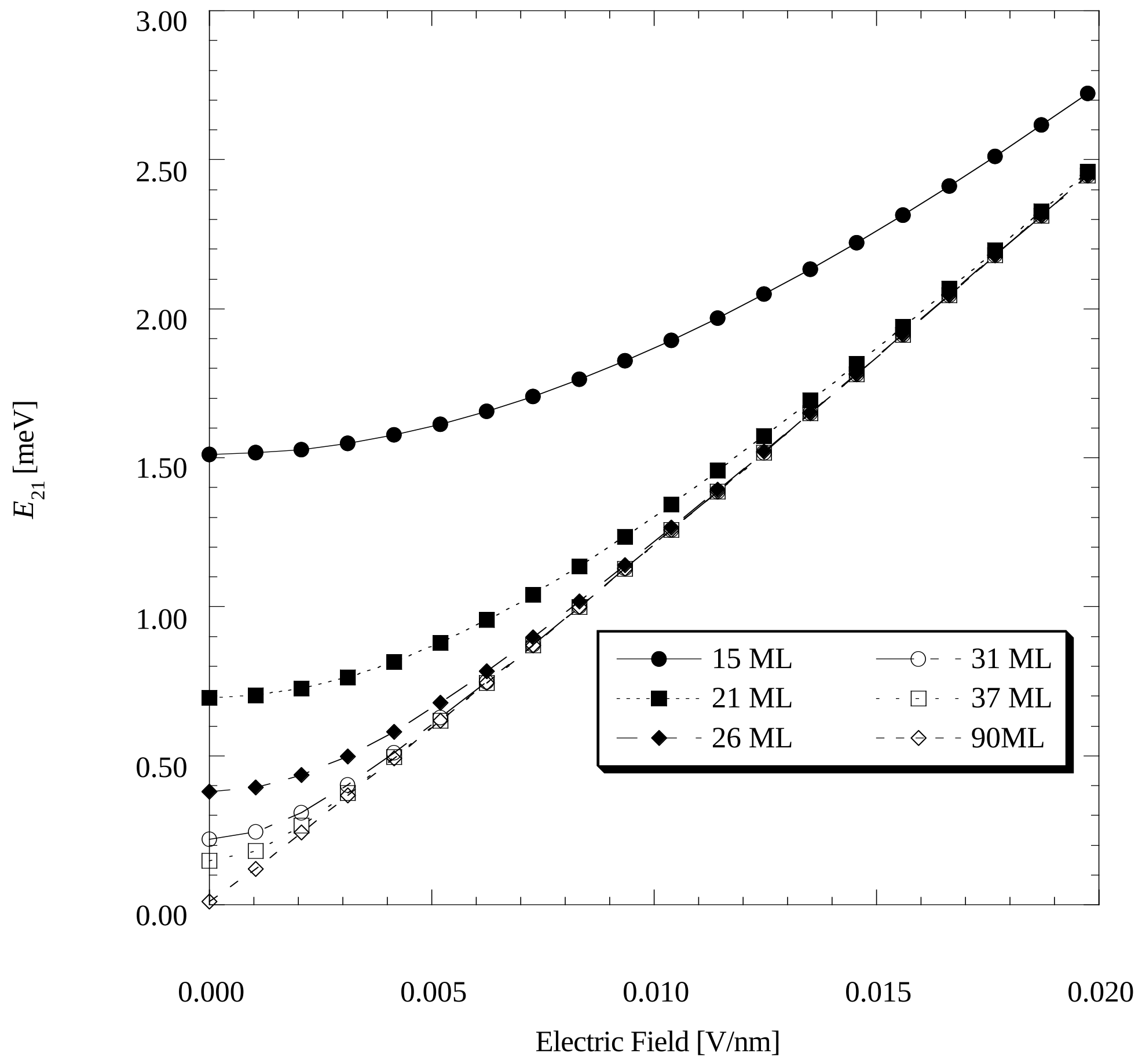

Figure 4